\documentclass{an}
\usepackage{graphicx}
\usepackage{times}
\Volume{0}               % issue in current year, format {NN}
\Year{2004}              % format: {YYYY}
\Pagespan{001}{005}      % format: {start page}{last page}

\begin{document}
\lhead[\thepage]{A.N. Author: Title}
\rhead[Astron. Nachr./AN~{\bf 325}, No. \volume\ (\yearofpublic)]{\thepage}
\headnote{Astron. Nachr./AN {\bf 325}, No. \volume, \pages\ (\yearofpublic) /
{\bf DOI} 10.1002/asna.\yearofpublic1XXXX}

\title{What can we learn from Accretion Disc Eclipse Mapping experiments?}

\author{Raymundo Baptista}
\institute{Departamento de F\'{\i}sica, UFSC, Campus Trindade, 88040-900,
			Florian\'opolis, Brazil}
\date{Received September 30; accepted November 14; published online {date}} 

\abstract{The accretion disc eclipse mapping method is an astrotomographic
inversion technique that makes use of the information contained in eclipse
light curves to probe the structure, the spectrum and the time evolution 
of accretion discs in cataclysmic variables. This paper presents examples
of eclipse mapping results that have been key to improve our
understanding of accretion physics.
}
\correspondence{bap@astro.ufsc.br}

\maketitle

\section{Introduction}

Cataclysmic Variables (CVs) are close interacting binaries in which mass 
is fed to a white dwarf (the primary) by a Roche lobe filling companion 
star (the secondary) via an accretion disc, which usually dominates the
ultraviolet and optical light of the system (Warner 1995).
Accretion discs in CVs cover a range of accretion rates, \.{M}, from
the low-mass transfer dwarf novae (the discs of which show recurrent 
outbursts on timescales of weeks to months) to the high-mass transfer
nova-like variables (the discs of which seems to be stuck more or less
in a steady state).

One of the difficulties in studying accretion disc physics comes from
the fact that the physical conditions in an accretion disc are expected 
to vary by large amounts with disc radius (the temperature distribution
in a steady-state disc decreases as $T\propto R^{-3/4}$), making the
integrated disc spectrum a complex combination of light emitted from 
regions of distinct physical conditions (e.g., Frank, King \& Raine 1992). 

In this regard, CVs are excellent sites for studying accretion physics
because their binary nature and relatively short orbital periods 
enable the application of powerful indirect imaging techniques.
These techniques overcome the intrinsic ambiguities associated with the
composite disc spectra by providing spatially resolved information about
accretion discs on angular scales of micro arcseconds -- well beyond 
the current direct imaging capabilities.

The Eclipse Mapping Method (Horne 1985) is one of these techniques.
It assembles the information contained in the shape of the eclipse into 
a map of the accretion disc surface brightness distribution.
The {\em eclipse map} is defined as a grid of intensities centred on the 
white dwarf and usually contained in the orbital plane.
The eclipse geometry is specified by the inclination $i$, the binary mass
ratio $q$ (=$M_2/M_1$, where $M_2$ and $M_1$ are the masses of, 
respectively, the secondary star and the white dwarf) and the phase of 
inferior conjunction.
Given the geometry, a model eclipse light curve can be calculated for any 
assumed brightness distribution in the eclipse map. A computer code then
iteratively adjusts the intensities in the map (treated as independent
parameters) to find the brightness distribution the model light curve
of which fits the data eclipse light curve within the uncertainties. 
The quality of the fit is checked with a consistency statistics,
usually $\chi^2$. Because the one-dimensional data light 
curve cannot fully constrain a two-dimensional map, additional freedom
remains to optimize some map property. A maximum entropy procedure
(e.g., Skilling \& Bryan 1984) is used to select, among all possible 
solutions, the one that maximizes the entropy of the eclipse map with 
respect to a smooth default map.
Details of the mathematical formulation of the problem can be found in 
Horne (1985) and Baptista (2001).
A movie ilustrating the iterations of an eclipse mapping experiment from 
start to convergence is available at \underbar{\bf 
www.astro.ufsc.br/$\sim$bap/slide1.gif}.

\section{A highlight of results} \label{high}

Early applications of the technique showed that accretion discs in 
outbursting dwarf novae (e.g, Horne \& Cook 1985) and in long-period 
novalike variables (e.g., Rutten, van Paradijs \& Tinbergen 1992) 
closely follow the $T \propto R^{-3/4}$ law expected a for steady-state 
disc, and that the radial temperature profile is essentially flat in
the short-period quiescent dwarf novae (e.g., Wood et~al 1989). 
This suggests that the viscosity in these short period systems is much 
lower in quiescence than in outburst, lending support to the disc 
instability model, and that their quiescent discs are far from being 
in a steady-state. 

From the $T(R)$ diagram it is possible to obtain an independent estimate
of the disc mass accretion rate, \.{M}.  
The diagram of \.{M} versus orbital period (e.g., Baptista 2001) suggests
a tendency among the steady-state discs of novalike variables to show 
larger \.{M} for longer binary period -- in agreement with current 
evolutionary scenarios for CVs  (Patterson 1984) -- and that the 
discs of novalike variables and outbursting dwarf novae of similar 
binary periods have comparable \.{M}. 
The mass accretion rates in the eclipse maps of novalike variables 
increase with disc radius. The departures from the steady-state disc 
model are more pronounced for the SW~Sex stars (period range 3-4~hs).
Illumination of the outer disc regions by the inner disc or mass ejection 
in a wind from the inner disc are possible explanations for this effect
(Rutten et~al 1992).

Multi-colour eclipse mapping is useful to probe the spectrum emitted by 
the different parts of the disc surface. Two-colour diagrams show that 
the inner disc regions of outbursting dwarf novae (e.g., Horne \& Cook 
1985) and of novalike variables (e.g., Horne \& Stiening 1985) are 
optically thick with a vertical temperature gradient less steep than 
that of a stellar atmosphere, and that optically thin, chromospheric 
emission appears to be important in the outer disc regions.
The fact that the emission from the inner disc regions is optically thick
thermal radiation opens the possibility to use a colour-magnitude diagram
to obtain independent estimates of the distance to the binary with a 
procedure similar to cluster main-sequence fitting.

\subsection{Spectral studies}

The eclipse mapping method advanced to the stage of delivering
spatially-resolved spectra of accretion discs with its application to
time-resolved eclipse spectrophotometry (Rutten et~al 1993).
The time-series of spectra is divided up into numerous spectral bins 
and light curves are extracted for each bin. The light curves are then
analyzed to produce a series of monochromatic eclipse maps covering the 
whole spectrum.  Finally, the maps are combined to obtain the spectrum 
for any region of interest on the disc surface.

\begin{figure}
\includegraphics[bb=1cm 1.5cm 16cm 21cm,scale=.44]{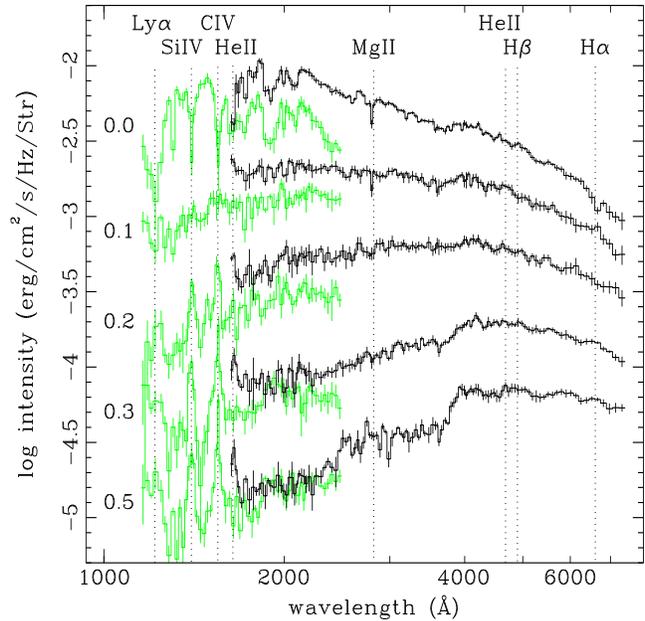}
\caption{ Spatially resolved spectra of the UX\,UMa accretion disc on
   August 1994 (gray) and November 1994 (black). The spectra were computed
   for a set of concentric annular sections (mean radius indicated on 
   the left in units of $R_{L1}$, the distance from disc centre to the 
   L1 point). The most prominent line transitions are indicated by 
   vertical dotted lines. From Baptista et~al (1998). }
\label{fig1}
\end{figure}

The spectral mapping analysis of nova-like variables (e.g., Baptista 
et~al 1998) shows that their inner accretion disc is characterized by 
a blue continuum filled with absorption bands and lines which cross 
over to emission with increasing disc radius (Fig.~\ref{fig1}). 
The continuum emission becomes progressively fainter and redder as one 
moves outwards, reflecting the radial temperature gradient.
These high-\.{M} discs seem hot and optically thick in their inner 
regions and cool and optically thin in their outer parts.
The spectrum of the infalling gas stream was found to be noticeably 
different from the disc spectrum at the same radius suggesting the 
existence of gas stream ``disc-skimming'' overflow that can be seen
down to the inner disc regions (e.g., Baptista et~al 2000b). 
%
%%%%%%%%%%%%%%%%%%%%%%%%%%%  FIGURE 1  %%%%%%%%%%%%%%%%%%%%%%%%%%%%%%
%%%%%%%%%%%%%%%%%%%%%%%%%%%%%%%%%%%%%%%%%%%%%%%%%%%%%%%%%%%%%%%%%%%%
%

The spectrum of the uneclipsed light in these nova-like systems 
shows strong emission lines and the Balmer jump in emission indicating 
an important contribution from optically thin gas. 
The lines and optically thin continuum emission are most probably
emitted in a vertically extended disc chromosphere + wind (e.g.,
Baptista et~al 2000b).
The uneclipsed spectrum of UX~UMa at long wavelengths is dominated by 
a late-type spectrum that matches the expected contribution from the 
secondary star (Rutten et~al 1994). Thus, the uneclipsed component 
provides an unexpected but interesting way of assessing the spectrum 
of the secondary star in CVs.

\subsection{Physical parameters studies} \label{ppem}

Vrielmann, Horne \& Hessman (2002a) developed a version of the
eclipse mapping method that simultaneously fits a set of multi-colour
light curves to directly map physical quantities in the accretion disc 
(in contrast to the classical mapping of disc surface brightnesses).
This {\em physical parameter} eclipse mapping requires the assumption of 
{\em a priori} spectral model for the disc emission relating the 
parameters to be mapped (e.g., temperature and surface density) to 
the observed surface brightness in a set of passbands.
The adopted spectral model is a pure hydrogen slab of gas in LTE
including only bound-free and free-free H and H$^-$ emission.
Although simple, this model is useful to distinguish between optically 
thin and optically thick disc regions and allows to estimate the disc 
surface density (for the optically thin regions).

The analysis of multicolor data of the dwarf novae HT~Cas and V2051~Oph 
with this technique provides evidence that their discs consist of a hot, 
optically thin chromosphere (responsible for the emission lines) on 
top of a cool, dense and optically thick disc layer (Vrielmann et~al. 
2002a, 2002b).

\subsection{Spatial studies} \label{spatial}

Eclipse mapping has also been a valuable tool to reveal that real
discs have more complex structures than in the simple axi-symmetric 
model.

Besides the normal outbursts, short-period dwarf novae (SU~UMa stars) 
exhibit superoutbursts in which superhumps develop with a period a few
per cent longer than the binary orbital period (e.g., Patterson 2001). 
O'Donoghue (1990) analyzed light curves of Z~Cha during superoutburst
with eclipse mapping techniques in order to investigate the location of 
the superhumps.  His analysis shows that the superhump light arises 
from the outer disc, and appears to be concentrated in the disc region 
closest to the secondary star.
This result helped to establish the superhump model of Whitehurst (1988), 
which proposed that they are the result of an increased tidal heating 
effect caused by the alignment of the secondary star and an slowly 
precessing eccentric disc.

Ultraviolet observations of the dwarf nova OY~Car in superoutburst
show dips in the light curve coincident in phase with the optical
superhump.  The eclipse mapping analysis of these data indicates the 
presence of an opaque disc rim, the thickness of which depends on the 
disc azimuth and is large enough for the rim to obscure the centre of 
the disc at the dip phase (Billington et~al 1996). 
These results are consistent with a model of superhumps as the 
consequence of time-dependent changes in the thickness of the edge of 
the disc, resulting in obscuration of the ultraviolet flux from the 
central regions and reprocessing of it into the optical part of the 
spectrum.

\begin{figure}
\begin{center}
\includegraphics[bb=1.5cm 4.8cm 14.5cm 20.7cm,angle=-90,scale=0.43]{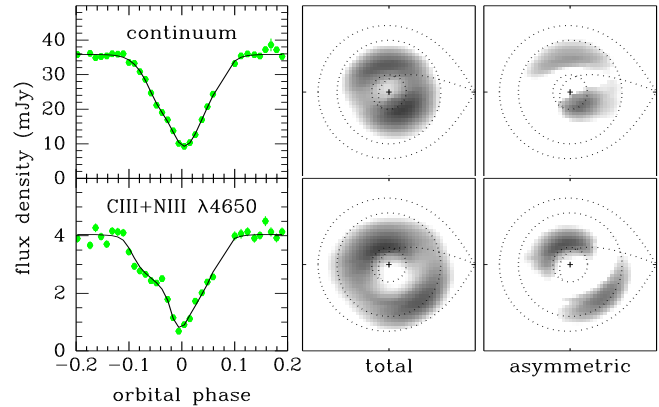}
\end{center}
\caption[]{ Eclipse mapping of spiral shocks in IP~Peg. The light curves
  are shown in the left-hand panels and the eclipse maps are displayed
  in the middle and right-hand panels in a logarithmic greyscale. Bright 
  regions are dark; faint regions are white. A cross marks the center of 
  the disc; dotted lines show the Roche lobe, the gas stream trajectory,
  and circles of radius 0.2 and $0.6\, R_{L1}$.
  The secondary is to the right of each map and the stars rotate
  counter-clockwise.  From Baptista et~al (2000a). }
\label{spiral}
\end{figure}

Tidally induced spiral shocks are expected to appear in dwarf novae 
discs during outburst as the disc expands and its outer parts feel more 
effectively the gravitational attraction of the secondary star.
Eclipse mapping of IP Peg during outburst (Baptista, Harlaftis \& 
Steeghs 2000a) helped to constrain the location and to investigate the 
spatial structure of the spiral shocks found in Doppler tomograms
(Steeghs, Harlaftis \& Horne 1997).
The spiral shocks are seen in the eclipse maps as two asymmetric arcs 
of $\sim 90$ degrees in azimuth extending from intermediate to the 
outer disc regions (Fig.~\ref{spiral}).
The comparison between the Doppler and eclipse maps reveal that the 
Keplerian velocities derived from the radial position of the shocks 
are systematically larger than those inferred from the Doppler tomography
indicating that the gas in the spiral shocks has sub-Keplerian velocities.
This experiment illustrates the power of combining the spatial information
obtained from eclipse mapping with the information on the disc dynamics
derived from Doppler tomography.
%
%%%%%%%%%%%%%%%%%%%%%%%%%%%  FIGURE 2  %%%%%%%%%%%%%%%%%%%%%%%%%%%%%%
%%%%%%%%%%%%%%%%%%%%%%%%%%%%%%%%%%%%%%%%%%%%%%%%%%%%%%%%%%%%%%%%%%%%

\subsection{Time-resolved studies}

Eclipse maps yield snapshots of an accretion disc at a given time. 
Time-resolved eclipse mapping may be used to track changes in the disc 
structure, e.g., to assess variations in mass accretion rate or to
follow the evolution of the surface brightness distribution through a 
dwarf nova outburst cycle.

Rutten et~al (1992b) obtained eclipse maps of the dwarf nova OY~Car 
along the rise to a normal outburst. Their maps show that the outburst 
starts in the outer disc regions with the development of a bright ring,
while the inner disc regions remain at constant brightness during the rise.
The flat radial temperature profile of quiescence and early rise changes,
within one day, into a steep distribution that matches a steady-state
disc model for \.{M}$= 10^{-9}\;M_\odot$\,yr$^{-1}$ at outburst maximum.
Their results suggest that an uneclipsed component develops during the
rise and contributes up to 15 per cent of the total light at 
outburst maximum. This may indicate the development of a 
vertically-extended (and largely uneclipsed) disc wind, or that the 
disc is flared during outburst.

Eclipse maps covering the full outburst cycle of the long-period dwarf 
nova EX~Dra (Baptista \& Catal\'an 2001) show the formation of a 
one-armed spiral structure in the disc at the early stages of the 
outburst and reveal how the disc expands during the rise until it fills 
most of the primary Roche lobe at maximum light. 
During the decline phase, the disc becomes progressively fainter until 
only a small bright region around the white dwarf is left at minimum light.
The evolution of the radial brightness distribution suggests the 
presence of an inward and an outward-moving heating wave during the 
rise and an inward-moving cooling wave in the decline.
The radial temperature distributions shows that, as a general 
trend, the mass accretion rate in the outer regions is larger than in 
the inner disc on the rising branch, while the opposite holds during 
the decline branch. Most of the disc appears to be in steady-state at 
outburst maximum and, interestingly, also during quiescence. 
It may be that the mass transfer rate in EX~Dra is sufficiently high to 
keep the inner disc regions in a permanent high viscosity, steady-state.
A movie with the sequence of eclipse maps of EX~Dra along its 
outburst cycle is available at \underbar{\bf 
www.astro.ufsc.br/$\sim$bap/slide2.gif}.

\subsection{Flickering mapping}

A long standing unsolved problem in accretion physics is related to 
the cause of flickering, the intrinsic brightness fluctuation of 0.1-1
magnitudes on timescales of seconds to minutes considered a basic 
signature of accretion.  Eclipse mapping of flickering light curves
(e.g., Welsh \& Wood 1995) is starting to shed light on this subject 
by spatially-resolving the flickering sources in CVs.

The orbital dependency of the flickering is obtained by measuring
the random variations due to flickering with respect to the mean flux
level in a set of light curves as a function of the orbital phase.
Baptista \& Bortoletto (2003) applied complementary techniques (e.g.,
Bruch 2000) to construct separate light curves of low- and 
high-frequency flickering of the dwarf nova V2051~Oph. Their eclipse 
mapping analysis reveals that the low-frequency flickering is 
associated to the bright spot and gas stream (suggesting that it is
caused by inhomogeneities in the mass transfer from the secondary star)
whereas the high-frequency flickering seems spread over the surface of
the disc with a radial distribution similar to that of the steady 
light (suggesting it reflects variability intrinsic to the disc
such as convection and/or magnetic turbulence).
%
%%%%%%%%%%%%%%%%%%%%%%%%%%%  FIGURE 3  %%%%%%%%%%%%%%%%%%%%%%%%%%%%%%
\begin{figure}
\begin{center}
\includegraphics[bb=1.3cm 4.1cm 19.2cm 21cm,angle=-90,scale=0.49]{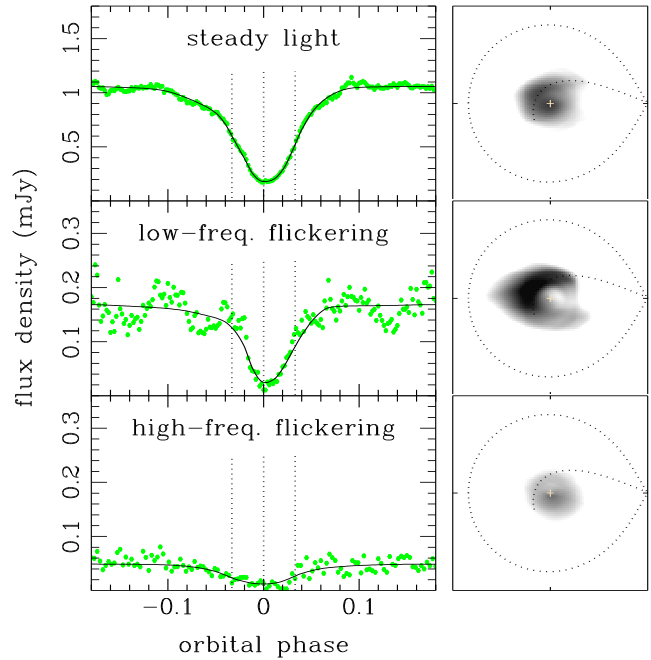}
\end{center}
\caption[]{ Flickering mapping of V2051~Oph. Top: steady-light. Middle:
  low-frequency flickering ($f<100$ mHz). Bottom: high-frequency 
  flickering ($f>6$ mHz). The notation is the same as in Fig.~\ref{spiral}. 
  Vertical dotted lines mark mid-eclipse and the ingress/egress phases 
  of the white dwarf.  The two flickering maps are in the same brightness
  scale.  From Baptista \& Bortoletto (2003). }
\label{fig4}
\end{figure}
%%%%%%%%%%%%%%%%%%%%%%%%%%%%%%%%%%%%%%%%%%%%%%%%%%%%%%%%%%%%%%%%%%%%

\section{Summary and future prospects}

Eclipse mapping is a powerful tool to probe the radial and vertical disc
structures, as well as to derive the physical conditions in accretion
discs. 
Partly thanks to the many experiments performed over the last two 
decades, our picture of accretion discs was enriched with an 
impressive set of new details such as gas outflow in disc winds, gas 
stream overflow, flared discs with azimuthal structure at their edge,
chromospheric disc line emission, ellipsoidal precessing discs, 
sub-Keplerian spiral shocks, and moving heating/cooling waves during 
disc outbursts.

Additional interesting eclipse mapping results are expected in the
near future. For example, fitting state-of-the-art disc atmosphere models 
to the spatially-resolved spectra is an obvious next step to the spectral
mapping experiments and will allow estimates of fundamental physical 
parameters of accretion discs, such as gas temperature, surface density, 
vertical temperature gradient, Mach number and viscosity, setting 
important additional constrains on current disc models.

\section*{Slide captions}

%
%%%%%%%%%%%%%%%%%%%%%%%%%%%  SLIDE 1  %%%%%%%%%%%%%%%%%%%%%%%%%%%%%%

\noindent {\bf Slide 1.} 
Left: Data (green dots with error bars) and model (solid line)
light curves.  A horizontal line depicts the uneclipsed component
to the total light.  Vertical dotted lines mark white dwarf ingress,
egress and mid-eclipse phases.  Labels indicate the number of 
interations and the value of $\chi^2$ in each case.
Right:  the corresponding eclipse map in a logarithmic greyscale
(dark regions are brighter).  The lower panel shows the asymmetric
part of the eclipse map. Dotted lines show the primary Roche lobe, 
the gas stream trajectory and a circle of radius $0.6\;R_{L1}$.
A cross mark the centre of the disc.
The slide shows the eclipse mapping run for nine iterations from 
close to start (niter=10) up to convergence (niter= 241).
The double-stepped, smooth asymmetric eclipse shape map into two
asymmetric arcs in the eclipse map (see also section 2.3).
\\

%
%%%%%%%%%%%%%%%%%%%%%%%%%%%  SLIDE 2  %%%%%%%%%%%%%%%%%%%%%%%%%%%%%%
\noindent {\bf Slide 2.} 
Sequence of light curves and eclipse maps of the dwarf nova EX Draconis
along its outburst cycle.  Top left: the radial intensity distributions
A dotted vertical line indicates the radial position of the bright spot 
in quiescence.  Large vertical ticks mark the position of the outer edge 
of the disc and short vertical ticks indicate the radial position of a 
reference intensity level.  Bottom left: The radial brightness temperature
distributions. Steady-state disc models for mass accretion rates of $\log$
\.{M}$= -7.5, -8.0, -8.5$, and $-9.0 \;M_\odot\;$yr$^{-1}$ are plotted as 
dotted lines for comparison.  The dot-dashed line marks the critical 
temperature above which the gas should remain in a steady, high-\.{M} 
regime.  The numbers in parenthesis indicate the time (in days) from the 
onset of the outburst.  Top right: sequence of light curves in quiescence 
(h), rise to maximum (a-b), during maximum light (c), through the decline 
phase (d-f), and at the end of the eruption (g), when the system goes 
through a low brightness state before recovering its quiescence level. 
Bottom right:  the corresponding eclipse maps in a logarithmic blackbody 
false color scale.  Dotted lines show the primary Roche lobe and the gas 
stream trajectory.  From Baptista \& Catal\'an (2001).

\end{document}